\documentclass[10pt,twocolumn,letterpaper]{article}

\usepackage[pagenumbers]{cvpr} 

\definecolor{cvprblue}{rgb}{0.21,0.49,0.74}
\usepackage[pagebackref,breaklinks,colorlinks,allcolors=cvprblue]{hyperref}

\def\paperID{SoS-4} 
\def\confName{CVPR}
\def\confYear{2026}

\usepackage{booktabs}
\usepackage{siunitx}
\usepackage{array,ragged2e}
\usepackage{tikz}
\usepackage[export]{adjustbox}
\usepackage{amsmath,amssymb}
\usepackage{multirow}

\newcolumntype{Y}{>{\RaggedRight\arraybackslash}p{0.55\linewidth}}

\title{PULSE: Privileged Knowledge Transfer from Rich to Deployable Sensors\\for Embodied Multi-Sensory Learning\thanks{Accepted at the CVPR 2026 Workshop on Sense of Space. This is the author's version of the work; the definitive version will appear in the IEEE/CVF CVPR 2026 Workshop Proceedings.}}

\author{
Zihan Zhao\textsuperscript{1} \quad
Kaushik Pendiyala\textsuperscript{2} \quad
Masood Mortazavi\textsuperscript{3} \quad
Ning Yan\textsuperscript{3}\\
\textsuperscript{1}University of California San Diego, San Diego, CA, USA\\
\textsuperscript{2}University of California Davis, Davis, CA, USA\\
\textsuperscript{3}IC Lab, Futurewei Technologies Inc., Santa Clara, CA, USA\\
{\tt\small \{ziz078@ucsd.edu, kpendiyala@ucdavis.edu, masood.mortazavi@futurewei.com, yan.ningyan@futurewei.com\}}
}

\begin{document}
\maketitle

\begin{abstract}
Multi-sensory systems for embodied intelligence, from wearable body-sensor networks to instrumented robotic platforms, routinely face a \emph{sensor-asymmetry} problem: the richest modality available during laboratory data collection is absent or impractical at deployment time due to cost, fragility, or interference with physical interaction.
We introduce \textbf{PULSE}, a general framework for \emph{privileged knowledge transfer} from an information-rich teacher sensor to a set of cheaper, deployment-ready student sensors.
Each student encoder produces \emph{shared} (modality-invariant) and \emph{private} (modality-specific) embeddings; the shared subspace is aligned across modalities and then matched to representations of a frozen teacher via multi-layer hidden-state and pooled-embedding distillation.
Private embeddings preserve modality-specific structure needed for self-supervised reconstruction, which we show is critical to prevent representational collapse.
We instantiate PULSE on the wearable stress-monitoring task, using electrodermal activity (EDA) as the privileged teacher and ECG, BVP, accelerometry, and temperature as students.
On the WESAD benchmark under leave-one-subject-out evaluation, PULSE achieves 0.994 AUROC and 0.988 AUPRC (0.965/0.955 on STRESS) \emph{without} EDA at inference, exceeding all no-EDA baselines and matching the performance of a full-sensor model that retains EDA at test time. We further demonstrate modality-agnostic transfer with ECG as teacher, provide extensive ablations on hidden-state matching depth, shared--private capacity, hinge-loss margin, fusion strategy, and modality dropout, and discuss how the framework generalizes to broader embodied sensing scenarios involving tactile, inertial, and bioelectrical modalities.
\end{abstract}

\section{Introduction}\label{sec:intro}

Multi-sensory systems for embodied intelligence routinely face a sensor-asymmetry problem: the richest modality available during data collection is absent or impractical at deployment.
This challenge is especially acute in wearable health monitoring.
Unlike camera-based systems, wearable physiological sensors enable continuous, unobtrusive monitoring of internal autonomic states in naturalistic settings: they operate during sleep, under clothing, and in private environments where visual sensing is infeasible or socially unacceptable.
Among wearable modalities, \emph{electrodermal activity} (EDA) occupies a privileged position: it is the only peripheral signal that directly indexes sympathetic nervous system activation via eccrine sweat gland innervation, making it the gold-standard physiological marker of acute stress~\cite{critchley2002brain,boucsein2012EDA,schmidt2018wesad,sanchezReolid2025scoping,roos2023wearable}.
However, EDA measurement requires Ag/AgCl electrodes with a constant-current source and is highly vulnerable to motion artifacts~\cite{hossain2022artifact}.
Consequently, most commercial wearables and longitudinal studies provide only ECG/PPG (BVP), inertial signals, and skin temperature; models trained with EDA simply cannot rely on it at test time.

This creates a dilemma: EDA carries privileged information about stress-related autonomic activation that cheaper modalities lack, yet discarding EDA entirely during training wastes a valuable supervisory signal.
The natural question is whether we can exploit EDA during training to improve the representations of deployment-grade sensors, without ever requiring EDA at inference.

\paragraph{Privileged knowledge transfer as a remedy.}
We address this through the Learning Using Privileged Information (LUPI) paradigm~\cite{vapnik2015lupi,lopezpaz2016unifying}: a teacher model trained on EDA transfers its learned representations into student encoders that consume only cheap, deployable modalities (ECG, BVP, ACC, TEMP).
Crucially, we \emph{freeze} the pretrained EDA teacher, providing a stable optimization target, rather than updating all encoders symmetrically as in missing-modality robustness methods~\cite{jiang2025physioomni,mordacq2024adapt,ibtehaz2024modally} that do not exploit the informational asymmetry between a privileged teacher and lighter-weight students.

\paragraph{The challenge: what to align, what to preserve.}
Na\"ively aligning all student representations to the teacher risks over-constraining the students: each modality carries unique, modality-specific structure (e.g., QRS morphology in ECG, pulse wave shape in BVP) that does not and should not match EDA.
Forcing full alignment destroys this structure and, as we show empirically, leads to representational collapse.
The key design insight is to separate each student's representation into a \emph{shared} subspace (modality-invariant content aligned to the teacher) and a \emph{private} subspace (modality-specific features preserved for reconstruction), and to transfer knowledge only through the shared channel.

\paragraph{Identified gap.}
LUPI and knowledge distillation have shown benefits in vision~\cite{lopezpaz2016unifying}, speech~\cite{markov2016lupi}, and affect modeling from video~\cite{makantasis2024labwild,aslam2024otpkd,aslam2024mtpkd,aslam2025diversitypkd}, while recent work distills PPG representations into accelerometer encoders at population scale~\cite{abbaspourazad2025accfm}.
EmotionKD~\cite{liu2023emotionkd} transfers across EEG and EDA for emotion recognition.
However, to our knowledge, no prior study combines (i) a frozen privileged-sensor teacher with (ii) an explicit shared--private embedding decomposition and (iii) multi-depth hidden-state alignment for wearable biosignal transfer, a design that disentangles transferable content from modality-specific structure at every layer of the network.

\paragraph{Contributions.}
We propose \textbf{PULSE}: \textbf{P}rivileged knowledge transfer \textbf{U}sing \textbf{L}ow-cost \textbf{SE}nsors for wearable stress monitoring:

\begin{enumerate}
    \item \textbf{EDA as a frozen teacher.} We pretrain an EDA encoder with self-supervised masked reconstruction, freeze it, and use its hidden-state embeddings at every layer plus a final pooled representation as privileged supervision. EDA is never required at inference.

    \item \textbf{Shared--private embedding decomposition.} Each student encoder splits its output into a shared subspace (aligned across modalities and matched to the teacher) and a private subspace (preserving modality-specific information for reconstruction). We show both components are essential: removing private capacity collapses representations, while removing shared capacity eliminates the transfer pathway.

    \item \textbf{Reconstruction as a collapse safeguard.} We demonstrate that knowledge distillation without an auxiliary reconstruction objective causes shared embeddings to collapse to a constant vector (cosine similarity $\approx 1.0$). The reconstruction branch acts as an information-preserving regularizer, a finding with implications for any multi-sensory distillation setup.

    \item \textbf{Comprehensive evaluation.} On WESAD~\cite{schmidt2018wesad} (15 subjects) and PhysioNet STRESS~\cite{hongn2025stress} (36 subjects), PULSE achieves 0.994 AUROC and 0.988 AUPRC (0.965/0.955 on STRESS) for binary stress detection \emph{without} EDA at inference, surpassing both a no-teacher baseline and a full-sensor model that retains EDA. We provide ablations to illustrate the effectiveness of each of our design choices.
\end{enumerate}

\paragraph{Broader relevance to multi-sensory embodied intelligence.}
While we instantiate PULSE on wearable physiological signals, the framework addresses a fundamental challenge shared across embodied sensing: how to compress information from a rich but impractical sensor into representations learned by cheaper, deployable modalities.
The shared--private decomposition, multi-depth distillation, and reconstruction-based collapse prevention are architecture-agnostic design principles applicable wherever sensor asymmetry arises, whether in tactile-to-IMU transfer for robotic manipulation, high-fidelity-to-consumer-grade transfer in XR sensing, or bioelectrical-to-optical transfer in wearable health monitoring.

\section{Related Work}\label{sec:relwork}

\paragraph{Learning using privileged information (LUPI).}
Vapnik and Izmailov~\cite{vapnik2015lupi} introduced LUPI, where a teacher has access to additional features unavailable to the student.
Lopez-Paz \etal~\cite{lopezpaz2016unifying} unified distillation and privileged information under a common framework, and Markov and Matsui~\cite{markov2016lupi} demonstrated gains in robust speech recognition.
In affect modeling, Makantasis \etal~\cite{makantasis2021privileged,makantasis2024labwild} transfer from laboratory physiological signals to video-only models for deployment in the wild.

\paragraph{Knowledge distillation in multi-sensory systems.}
EmotionKD~\cite{liu2023emotionkd} distills across heterogeneous biosignals (EEG and EDA) for single-modality emotion recognition at test time.
Aslam \etal~\cite{aslam2024otpkd,aslam2024mtpkd,aslam2025diversitypkd} develop optimal-transport and multi-teacher distillation for expression recognition.
Abbaspourazad \etal~\cite{abbaspourazad2025accfm} distill PPG representations into accelerometer encoders at population scale, producing generalist wearable foundation models.

\paragraph{Missing-modality robustness.}
PhysioOmni~\cite{jiang2025physioomni} and ADAPT~\cite{mordacq2024adapt} align modalities symmetrically to handle missing channels, while Ibtehaz and Mortazavi~\cite{ibtehaz2024modally} pursue simultaneous alignment and reconstruction for multi-lead ECG.
These approaches treat all modalities as peers rather than exploiting asymmetry between a privileged teacher and deployment students.

\paragraph{Self-supervised pretraining for time series.}
Several self-supervised objectives have been proposed for time-series representation learning.
Contrastive Multi-Segment Coding (CMSC)~\cite{kiyasseh2021clocs} treats adjacent non-overlapping segments of the same recording as positive pairs and applies contrastive learning to capture temporal invariance.
CLIP-style approaches~\cite{radford2021clip} align representations across paired views (e.g., time and frequency domains, or different sensor channels) with an InfoNCE objective and have been adapted to biosignals~\cite{zhang2022tfc}.
TS-TCC~\cite{eldele2021tstcc} combines temporal and contextual contrasting with strong/weak augmentations.
Masked autoencoders (MAE)~\cite{he2022mae}, originally developed for vision, reconstruct randomly masked patches and have been adapted to physiological signals~\cite{liu2024biofame}.

\paragraph{Positioning.}
Unlike symmetric alignment methods, PULSE uses a \emph{frozen} privileged-modality teacher that provides a stable optimization target at every encoder layer.
Unlike prior PKD for affect~\cite{aslam2024otpkd,liu2023emotionkd}, we introduce an explicit shared--private decomposition that prevents over-constraining modality-specific features, and we demonstrate that reconstruction loss is essential for preventing collapse during distillation, a finding with implications for any multi-sensory transfer setup.

\section{Datasets and Preprocessing}\label{sec:data}

\paragraph{Source.}
We use the publicly available \textbf{WESAD} benchmark~\cite{schmidt2018wesad}, which provides synchronized physiological signals from 15 subjects wearing a RespiBAN chest band and an Empatica~E4 wristband during a 35-min protocol (baseline $\rightarrow$ social-stress $\rightarrow$ comedy).
Signals recorded on the wrist include \emph{electrodermal activity} (EDA, 4\,Hz), \emph{photoplethysmography} (BVP, 64\,Hz), skin \emph{temperature} (TEMP, 4\,Hz), and tri-axial \emph{acceleration} (ACC, 32\,Hz); the chest strap provides \emph{ECG}, respiration, EMG, and ACC at 700\,Hz.
This multi-sensor setup exemplifies the sensor-asymmetry problem: EDA is the gold-standard stress indicator but requires specialized electrodes, while the remaining modalities are available on commodity wearables.

\paragraph{Resampling.}
All channels are resampled to a uniform 64\,Hz target rate: first-order polyphase filtering for high-frequency signals (ECG, BVP, ACC), linear interpolation for low-frequency channels (EDA, TEMP).

\paragraph{Signal cleaning.}
\begin{itemize}
\setlength\itemsep{0.2em}
\item \textbf{EDA}: tonic drift removed with a first-order $0.05$\,Hz high-pass; phasic band retained with a $0.05$--$1$\,Hz Butterworth filter.
\item \textbf{BVP}: band-pass $0.5$--$2$\,Hz to isolate pulse wave.
\item \textbf{ECG}: band-pass $0.5$--$40$\,Hz.
\item \textbf{Motion}: net acceleration computed as $\lVert \mathrm{ACC}_{x,y,z} \rVert$ from the \emph{wrist} accelerometer only; chest-band ACC is excluded to match wrist-only deployment hardware.
\end{itemize}

\paragraph{Segmentation.}
Cleaned, synchronized streams are cut into 60-second windows with a 0.25-second stride (96\% overlap), following~\cite{schmidt2018wesad,strzinar2023freqEDA}.
Label~0 (\textit{transient}) samples are discarded; only windows whose entire 60\,s lie in a single class $\{1{=}\text{baseline}, 2{=}\text{stress}, 3{=}\text{amusement}\}$ are kept.
Windows whose ECG or BVP standard deviation is lower than $0.02$ (after $z$-scoring) are rejected to remove sensor drop-outs.

\paragraph{Normalization.}
Per-subject $z$-score parameters are estimated from all baseline ($\text{label}{=}1$) windows and applied to every channel, preserving inter-subject differences while removing long-term drift.

\paragraph{Leave-one-subject-out (LOSO) folds.}
The pipeline yields ${\sim}8 \times 10^3$ valid windows per subject (3,840 samples per window).
We build fifteen LOSO folds: in fold~$k$, subject $S_k$ is the test set and the remaining 14 subjects form the training set.

\subsection{Second Dataset: PhysioNet STRESS}\label{subsec:stress_data}

To evaluate cross-dataset generalization, we additionally use the \textbf{Wearable Device Dataset from Induced Stress and Structured Exercise Sessions}~\cite{hongn2025stress}, a publicly available dataset on PhysioNet containing physiological recordings from 36 healthy volunteers. Data was collected using the Empatica~E4 wristband during a structured acute stress induction protocol involving mathematical and emotional tasks. The device records \emph{blood volume pulse} (BVP, 64\,Hz), \emph{electrodermal activity} (EDA, 4\,Hz), \emph{skin temperature} (TEMP, 4\,Hz), and tri-axial \emph{acceleration} (ACC, 32\,Hz). Unlike WESAD, no chest-band ECG is available; this makes the dataset a natural testbed for the wrist-only deployment scenario.

We apply the same preprocessing pipeline as WESAD (resampling to 64\,Hz, signal cleaning, 60\,s windowing, per-subject baseline normalization) to the PhysioNet STRESS data, using the rest/baseline segments for normalization.

\section{Methods}\label{sec:methods}

\begin{figure*}[t]
    \centering
    \includegraphics[width=0.8\linewidth, trim=0 20pt 0 60pt, clip]{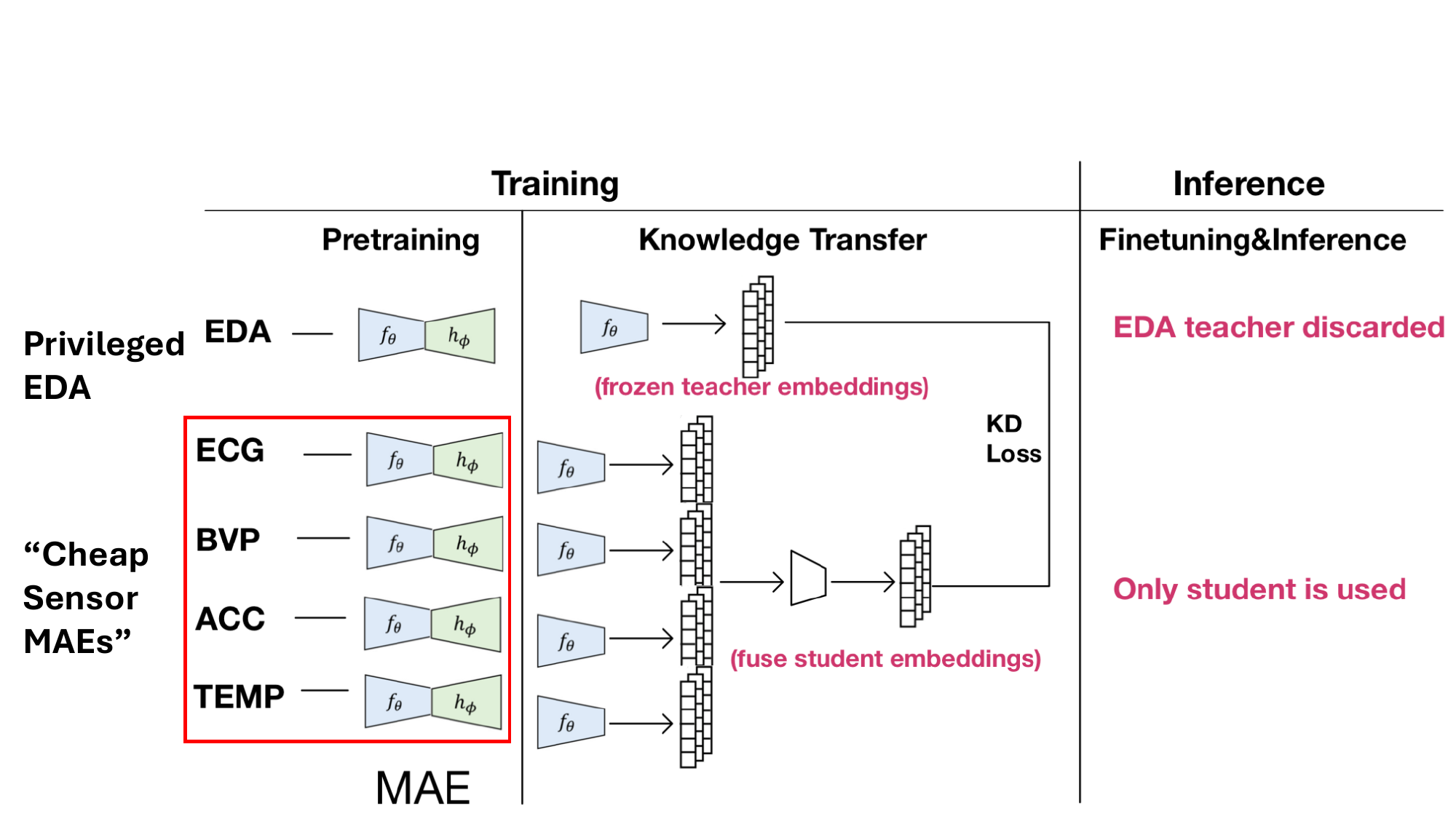}
    \caption{\textbf{The PULSE Framework.} Our framework uses privileged knowledge transfer from a frozen teacher encoder (here: EDA) to students built on low-cost, deployment-ready sensors. In the pretraining stage, student encoders learn modality-invariant \emph{shared} embeddings alongside modality-specific \emph{private} embeddings. Knowledge transfer aligns the students' shared embeddings with the privileged teacher. During finetuning, only the learned student embeddings are used for supervised classification; no privileged sensor is required at inference.}
    \label{fig:overview}
\end{figure*}

Our training pipeline, summarized in \cref{fig:overview}, proceeds in three stages: (1) self-supervised pretraining of per-modality masked autoencoders with cross-modal alignment, (2) privileged knowledge transfer from a frozen teacher, and (3) supervised finetuning with only deployment sensors. No privileged sensor data is required at inference.

\begin{figure*}[t]
    \centering
    \includegraphics[width=0.7\linewidth]{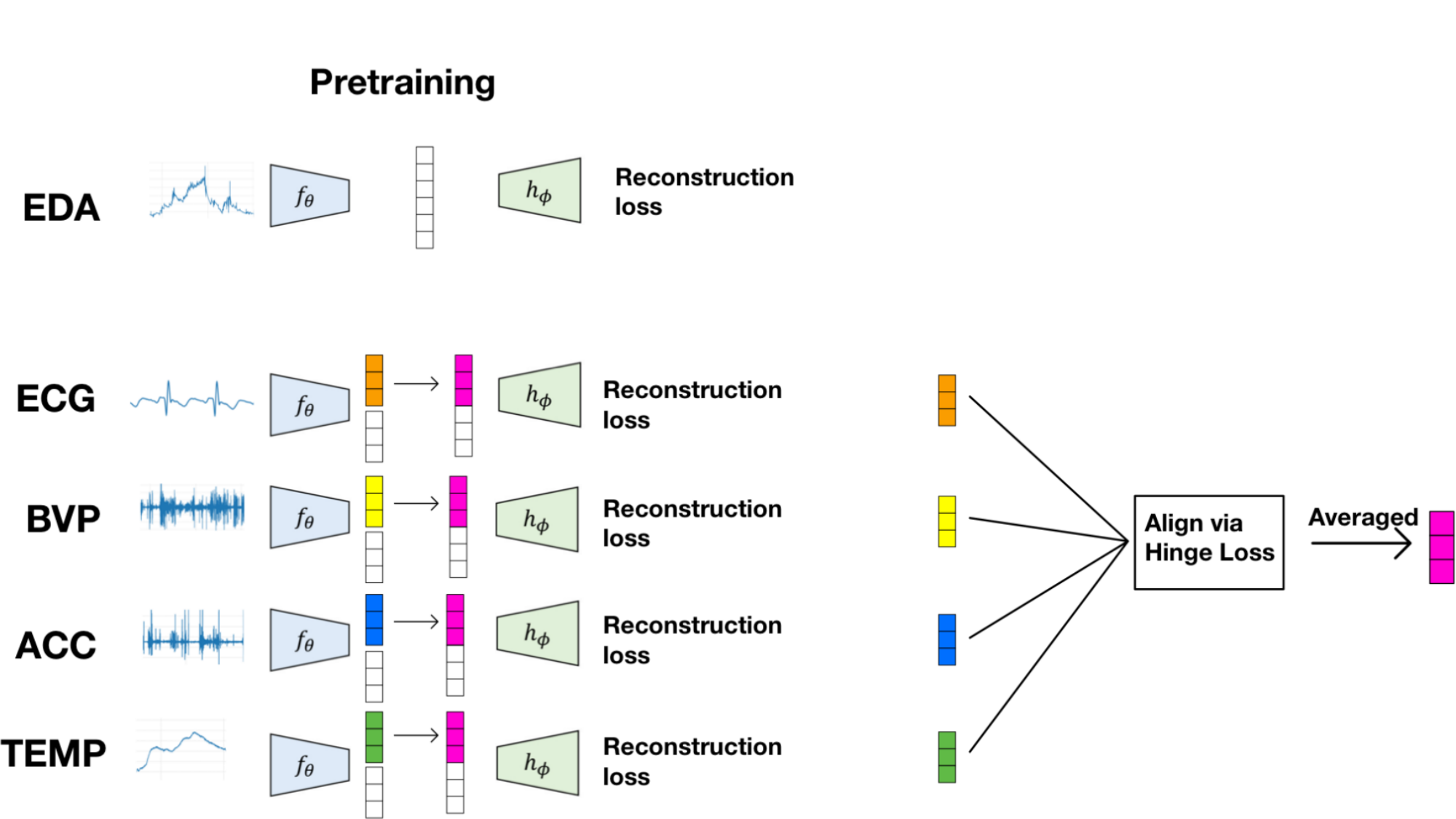}
    \caption{\textbf{Pretraining Setup.} Each student encoder outputs shared embeddings (colored) and private embeddings (white). Shared embeddings are aligned across modalities via a hinge loss, then averaged into a single modality-invariant embedding (magenta). This averaged embedding, together with private embeddings, drives reconstruction. The teacher MAE is pretrained separately via reconstruction loss only.}
    \label{fig:pretraining}
\end{figure*}

\begin{figure*}[t]
    \centering
    \includegraphics[width=0.9\linewidth, trim=0 200pt 0 60pt, clip]{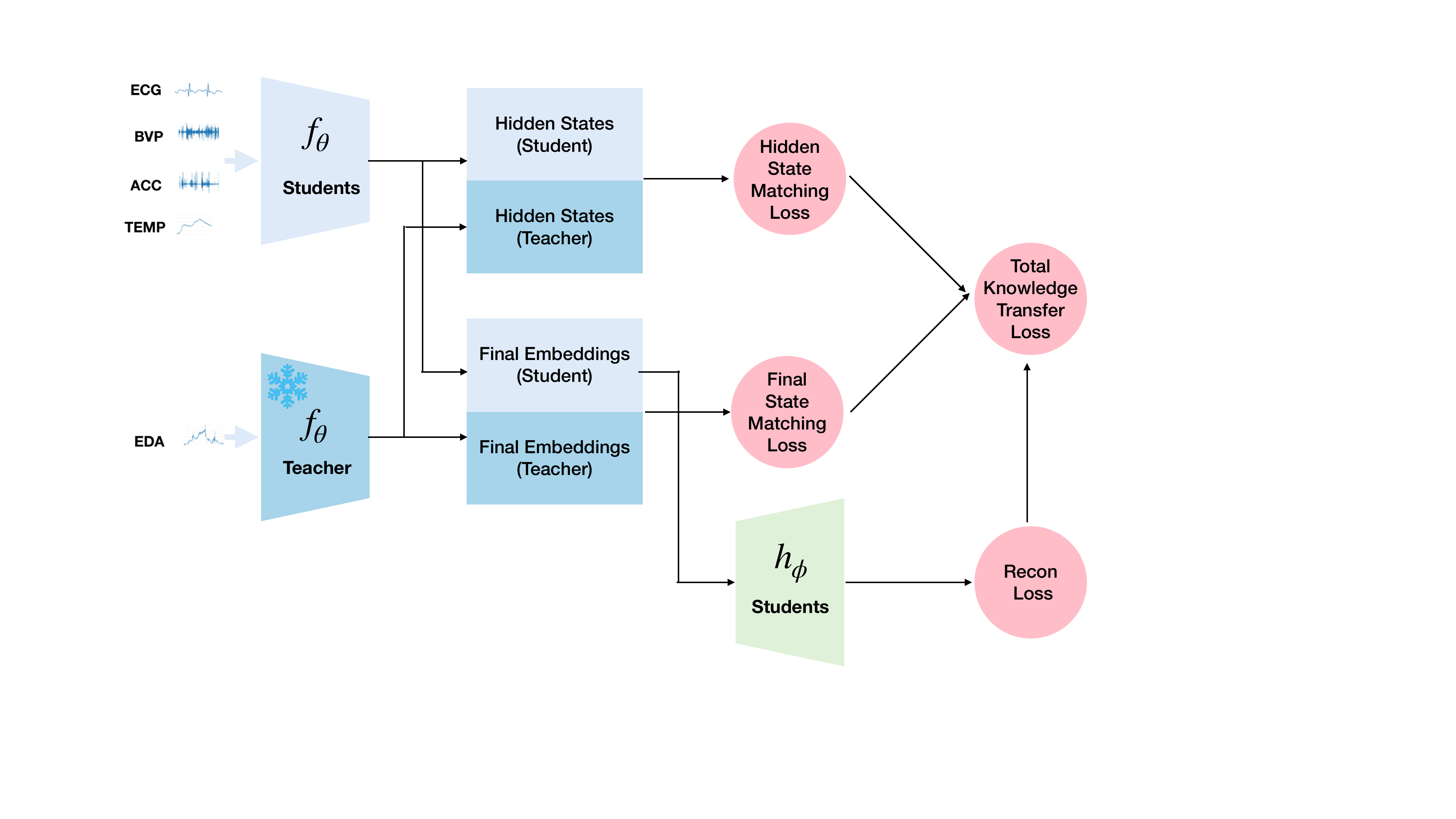}
    \caption{\textbf{Knowledge Transfer.} The frozen EDA teacher encoder and the student encoders (ECG, BVP, ACC, TEMP) process their respective inputs in parallel. Knowledge is transferred through two complementary objectives: (1) a hidden-state matching loss that aligns fused student hidden representations with the teacher's intermediate representations at every encoder layer, and (2) a final-state matching loss that matches the pooled last-layer embeddings. An auxiliary reconstruction loss from the student decoders preserves modality-specific information and prevents representational collapse. The three terms are combined into the total knowledge transfer loss. The teacher encoder remains frozen throughout; only student parameters are updated.}
    \label{fig:kd_diagram}
\end{figure*}

\subsection{Pretraining}\label{subsec:pretrain}

Before knowledge transfer, the student \texttt{PhysioMAE} models (ECG, BVP, ACC, TEMP) are pretrained jointly in a self-supervised manner, while the teacher \texttt{PhysioMAE} (EDA) is pretrained separately (\cref{fig:pretraining}). Inputs of length 3,840 samples at 64\,Hz are split into non-overlapping patches (size 96), embedded, and processed by a transformer encoder (embedding dimension 1024, depth 8, heads 8). A lightweight decoder (dimension 512, depth 4, heads 8) reconstructs masked signal patches.

\paragraph{Shared--private decomposition.}
Each student encoder outputs two types of embeddings: \emph{shared} embeddings (modality-invariant) and \emph{private} embeddings (modality-specific), separated via a random binary mask. The teacher encoder does not undergo the shared/private split, since it does not participate in cross-modal alignment.

\paragraph{Alignment loss.}
Shared embeddings are aligned using a hinge loss with in-batch negatives. Let $\mathcal{P}$ denote positive (matched) cross-modal pairs. For a positive pair $(i,j) \in \mathcal{P}$ with shared embeddings $s_i, s_j$ and in-batch negatives $\mathcal{N}(j), \mathcal{N}(i)$:
\small
\paragraph{Alignment loss.}
Shared embeddings are aligned using a hinge loss with in-batch negatives. Let $\mathcal{P}$ denote positive cross-modal pairs, and let $\alpha=0.2$ be the margin. Then
\begin{equation}
\small
\begin{aligned}
\mathcal{L}_{\mathrm{align}}
= \frac{1}{|\mathcal P|}\sum_{(i,j)\in\mathcal P}\Bigg[
&\frac{1}{|\mathcal N(j)|}\sum_{j'\in\mathcal N(j)}
\max\Big(0,\,
\cos(s_i,s_{j'})  \\
&\qquad\qquad\qquad\qquad
-\cos(s_i,s_j)+\alpha\Big) \\
&+\frac{1}{|\mathcal N(i)|}\sum_{i'\in\mathcal N(i)}
\max\Big(0,\,
\cos(s_{i'},s_j) \\
&\qquad\qquad\qquad\qquad
-\cos(s_i,s_j)+\alpha\Big)
\Bigg].
\end{aligned}
\end{equation}
where $\alpha{=}0.2$ is the margin.

\paragraph{Reconstruction loss.}
The averaged shared embedding and private embeddings are decoded to reconstruct masked patches $\Omega$:
\begin{equation}
\small
\mathcal{L}_{\text{rec}} = \frac{1}{|\Omega|}\sum_{\omega\in\Omega} \| x_\omega - \hat{x}_\omega \|^2 .
\end{equation}

\paragraph{Total pretraining loss.}
$\mathcal{L}_{\text{pre}} = \lambda_{\text{align}}\mathcal{L}_{\text{align}} + \lambda_{\text{rec}}\mathcal{L}_{\text{rec}}$, with $\lambda_{\text{align}}{=}\lambda_{\text{rec}}{=}1$. We pretrain for 300 epochs with Adam ($lr{=}10^{-4}$, batch size 128).

\subsection{Knowledge Transfer}\label{subsec:kt}

We freeze the teacher encoder and distill it into the four student encoders (\cref{fig:kd_diagram}). All models share the same ViT-style configuration.

\paragraph{Transfer heads and fusion.}
A lightweight transfer head (i) layer-normalizes student embeddings, (ii) projects into shared/private subspaces via per-modality linear layers, and (iii) fuses shared embeddings across modalities with a linear fusion layer initialized to exact averaging.

\paragraph{What is transferred.}
We align (a) \emph{hidden tokens} at all student layers against corresponding teacher layers and (b) a \emph{final pooled embedding} from the last layer. No logits or labels are used.

For each matched layer $\ell$:
\begin{equation}
\small
\mathcal{L}_{\text{hid}} = \frac{1}{|\mathcal{L}|}\sum_{\ell\in\mathcal{L}}
\Big(1-\cos\!\big\langle \mathtt{Fuse}(\{S_m^\ell\}), T^{\ell} \big\rangle\Big),
\end{equation}
and for the final embedding:
\begin{equation}
\small
\mathcal{L}_{\text{emb}} = 1-\cos\!\big\langle \mathrm{mean}_t\, \mathtt{Fuse}(\{S_m^{\mathrm{final}}\}), \mathrm{mean}_t\, T^{\mathrm{final}} \big\rangle .
\end{equation}

\paragraph{Optional regularizers.}
Two optional terms can be toggled: (i) a decorrelation penalty $\mathcal{L}_{\text{perp}}$ between shared and private embeddings, and (ii) a reconstruction loss from each student's MAE decoder (masking ratio 0.5). We find reconstruction essential for avoiding collapse (\cref{sec:discussion}).

\paragraph{Total loss.}
$\mathcal{L} = \lambda_{\text{hid}}\mathcal{L}_{\text{hid}} + \lambda_{\text{emb}}\mathcal{L}_{\text{emb}} + \lambda_{\text{rec}}\mathcal{L}_{\text{rec}} + \lambda_{\text{perp}}\mathcal{L}_{\text{perp}}$,
with defaults $\lambda_{\text{hid}}{=}\lambda_{\text{emb}}{=}1$, $\lambda_{\text{rec}}{=}0.1$, $\lambda_{\text{perp}}{=}0$. We train for 100 epochs with Adam ($lr{=}10^{-4}$, batch size 128).

\subsection{Finetuning and Inference}\label{subsec:finetune}

At test time, the privileged modality is absent. Student encoders consume only deployment-grade sensors, form a fused shared representation, and classify via a lightweight 2-layer MLP (hidden dimension 4). Encoders remain frozen; only the MLP head is trained.

\paragraph{Overfitting control.}
We uniformly subsample training windows by $1/40$ to reduce correlation between overlapping windows and use a deliberately small classification head to limit capacity, improving cross-subject generalization.

\paragraph{Optimization.}
We train each fold for 300 epochs with Adam ($lr{=}10^{-3}$), batch size 128, cosine LR schedule. Model selection uses highest validation AUPRC.

\section{Experiments and Results}\label{sec:experiments}

We evaluate five training configurations summarized in \cref{tab:setups} and report aggregate metrics in \cref{tab:finetune}.

\begin{table*}[t]
\centering
\small
\setlength{\tabcolsep}{4pt}
\renewcommand{\arraystretch}{1.1}
\caption{Training and evaluation configurations. ``Cheap sensors'' refers to deployment-grade modalities (ECG, BVP, ACC, TEMP). The privileged sensor (EDA) is used only during training.}
\label{tab:setups}
\begin{tabular}{clp{4.2cm}cp{2.8cm}p{4.2cm}}
\toprule
\textbf{ID} & \textbf{Name} & \textbf{Train-time inputs} & \textbf{Teacher?} & \textbf{Test-time inputs} & \textbf{Purpose} \\
\midrule
A & No-teacher baseline & Cheap sensors only & -- & Cheap sensors only & Achievable without privileged data \\
B & Symmetric alignment & Cheap sensors + EDA (all updated jointly) & -- & Cheap sensors only & Standard cross-modal alignment \\
C & \textbf{PULSE} & Cheap sensors + EDA (EDA frozen $\rightarrow$ students match) & Yes & Cheap sensors only & Privileged knowledge transfer \\
D & Full-sensor baseline & Same as B & -- & Cheap sensors + EDA & Upper bound if privileged sensor is kept \\
E & Teacher only (EDA) & Pretrain EDA MAE; finetune classifier & -- & EDA only & Teacher quality verification \\
\bottomrule
\end{tabular}
\end{table*}

\begin{table*}[t]
\centering
\small
\setlength{\tabcolsep}{8pt}
\renewcommand{\arraystretch}{1.1}
\caption{Binary stress classification (baseline vs.\ stress) under LOSO. Mean $\pm$ std across 15 subjects. PULSE uses no privileged sensor at inference.}
\label{tab:finetune}
\begin{tabular}{llccc}
\toprule
\textbf{ID / Name} & \textbf{Test-time inputs} & \textbf{AUROC} & \textbf{AUPRC} & \textbf{Accuracy (\%)} \\
\midrule
A / No-teacher baseline & Cheap sensors & $0.963 \pm 0.050$ & $0.937 \pm 0.101$ & $91.64 \pm 6.61$ \\
B / Symmetric alignment & Cheap sensors & $0.972 \pm 0.031$ & $0.944 \pm 0.061$ & $88.83 \pm 6.24$ \\
C / \textbf{PULSE} & Cheap sensors & $\mathbf{0.994 \pm 0.011}$ & $\mathbf{0.988 \pm 0.022}$ & $\mathbf{96.08 \pm 4.52}$ \\
D / Full-sensor baseline & Cheap sensors + EDA & $0.983 \pm 0.028$ & $0.963 \pm 0.048$ & $90.74 \pm 5.58$ \\
E / Teacher only (EDA) & EDA & $0.962 \pm 0.067$ & $0.924 \pm 0.122$ & $87.20 \pm 17.69$ \\
\bottomrule
\end{tabular}
\end{table*}

\paragraph{Metric choice.}
We report threshold-independent AUROC and AUPRC as primary metrics, avoiding the instability of threshold-dependent accuracy under cross-subject evaluation. Accuracy at a fixed threshold is included as a secondary summary.

\paragraph{Main observations.}
Symmetric alignment (B) yields only a minor improvement over the no-teacher baseline (A) when the privileged modality is removed at test time: enforcing cross-modal consistency alone does not transfer enough privileged information.
PULSE (C) delivers consistent gains across all metrics.
The frozen teacher provides a stable optimization target that shapes the students' shared representations toward the teacher's signal structure, and this benefit persists when only cheap sensors are used at inference.

Notably, PULSE (C) \emph{matches or exceeds} the full-sensor model (D) on AUROC/AUPRC, despite not using EDA at test time.
This counter-intuitive result does not imply that EDA is uninformative at inference; rather, it reflects an optimization advantage of the frozen-teacher design.
In configuration~D, all five encoders (including EDA) are jointly optimized end-to-end through a 2-layer MLP with hidden dimension~4.
With only 15 LOSO subjects, this joint optimization is prone to overfitting to subject-specific EDA artifacts (e.g., tonic drift residuals or motion-contaminated phasic bursts that pass preprocessing), causing the full-sensor model to underperform on held-out subjects.
By contrast, PULSE freezes the EDA encoder after self-supervised pretraining, preventing it from co-adapting with the students, and transfers only the generalizable structure of EDA representations into the shared subspace.
The frozen teacher thus acts as a data-dependent regularizer: it anchors student optimization to a stable target learned from the full training population, effectively denoising and compressing the privileged signal.
This effect is most visible in the three-class setting (\cref{tab:tri_class,fig:per_fold_3class}), where PULSE beats the full-sensor baseline by a large margin (0.956 vs. 0.812 AUROC) and where concrete per-subject evidence supports the regularizer interpretation; see \cref{subsec:three_class} for analysis.

\subsection{Three-Class Classification}\label{subsec:three_class}

\begin{table*}[!ht]
\centering
\small
\caption{Three-class classification (baseline / stress / amusement) under LOSO. PULSE uses no EDA at inference.}
\label{tab:tri_class}
\begin{tabular}{lccc}
\toprule
Model & AUROC & AUPRC & Acc.\ (\%) \\
\midrule
no-EDA baseline        & $0.891 \pm 0.090$ & $0.810 \pm 0.130$ & $71.62 \pm 10.27$ \\
symmetric alignment    & $0.811 \pm 0.103$ & $0.710 \pm 0.126$ & $69.12 \pm 8.82$ \\
\textbf{PULSE}         & $\mathbf{0.956 \pm 0.058}$ & $\mathbf{0.894 \pm 0.115}$ & $\mathbf{85.40 \pm 7.18}$ \\
full-sensor baseline   & $0.812 \pm 0.128$ & $0.703 \pm 0.124$ & $73.38 \pm 7.89$ \\
\bottomrule
\end{tabular}
\end{table*}

The gap between PULSE and all baselines \emph{widens} in the three-class setting (\cref{tab:tri_class}), suggesting that fine-grained distinctions between affective states benefit disproportionately from privileged supervision.

\begin{figure*}[t]
\centering
\includegraphics[width=0.85\linewidth]{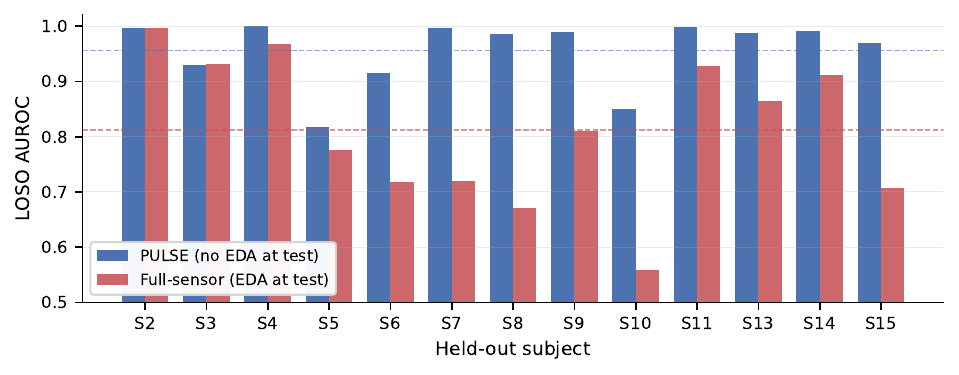}
\caption{\textbf{Per-subject LOSO AUROC on three-class classification.} Dashed lines indicate mean AUROC (blue: PULSE, red: full-sensor baseline). PULSE outperforms the full-sensor model on 11 of 13 folds, with the two remaining folds (S2, S3) within 0.002 AUROC. The full-sensor model suffers catastrophic drops on S6--S10 and S15, while PULSE maintains stable AUROC across all subjects. The across-fold standard deviation is more than halved under PULSE (0.060 vs.\ 0.133), consistent with the frozen EDA teacher acting as a variance-reducing regularizer.}
\label{fig:per_fold_3class}
\end{figure*}

\paragraph{Per-subject analysis.}
\cref{fig:per_fold_3class} breaks down \cref{tab:tri_class} by held-out subject, providing concrete evidence for the regularizer interpretation of the frozen teacher.
PULSE outperforms the full-sensor baseline on 11 of 13 LOSO folds\footnote{Subject~S1 is absent from WESAD and S12 is used as a fixed validation fold across all runs, leaving 13 LOSO test subjects.}, with the two remaining folds (S2, S3) essentially tied (gap $<0.002$ AUROC).
More revealing is the \emph{shape} of the per-fold distribution: PULSE maintains AUROC above 0.85 for every subject except S5 (0.82), while the full-sensor model suffers catastrophic drops on S10 (0.56), S8 (0.67), S15 (0.71), S6 (0.72), and S7 (0.72).
The across-fold AUROC standard deviation is more than halved under PULSE (0.060 vs.\ 0.133 for full-sensor), a classic signature of reduced variance that a regularizer should produce.
These catastrophic per-subject failures in the full-sensor model are consistent with our interpretation that joint end-to-end optimization of all five encoders through a small classification head overfits to subject-specific EDA artifacts when those artifacts happen to correlate spuriously with training-subject labels; when such a subject is held out for test, the overfit EDA pathway provides no useful signal and can actively harm the prediction.
PULSE sidesteps this failure mode by freezing the EDA encoder, so students can only access EDA information through its stable pretrained representations rather than through potentially artifact-coupled gradients.

\subsection{Results on PhysioNet STRESS}\label{subsec:stress_results}

We evaluate PULSE on the PhysioNet STRESS dataset~\cite{hongn2025stress} to test generalization beyond WESAD. Since this dataset provides only wrist-worn Empatica~E4 signals (BVP, EDA, ACC, TEMP) without chest-band ECG, the student set consists of BVP, ACC, and TEMP, with EDA as the privileged teacher. Results under LOSO across 36 subjects are in \cref{tab:stress_results}.

\begin{table*}[!ht]
\centering
\small
\caption{Binary stress classification on PhysioNet STRESS~\cite{hongn2025stress} under LOSO (36 subjects). Student modalities: BVP, ACC, TEMP. EDA is privileged (training only).}
\label{tab:stress_results}
\begin{tabular}{lccc}
\toprule
Model & AUROC & AUPRC & Acc.\ (\%) \\
\midrule
No-teacher baseline  & $0.827 \pm 0.117$ & $0.716 \pm 0.168$ & $82.61 \pm 11.2$ \\
Symmetric alignment  & $0.734 \pm 0.196$ & $0.678 \pm 0.191$ & $78.57 \pm 13.74$ \\
\textbf{PULSE}       &  $\mathbf{0.965 \pm 0.045}$ & $\mathbf{0.955 \pm 0.045}$ & $\mathbf{88.88 \pm 6.93}$ \\
Full-sensor baseline & $0.835 \pm 0.135$ & $0.722 \pm 0.141$ & $76.98 \pm 11.44$  \\
\bottomrule
\end{tabular}
\end{table*}

\section{Ablation Studies}\label{sec:ablations}

We present three ablations that validate the core design principles of PULSE. Additional hyperparameter sweeps and secondary ablations are in the appendix (\cref{app:ablation}).

\subsection{Hidden-State Matching Depth}\label{subsec:hid_match}

\begin{table}[t]
\centering
\small
\caption{Effect of hidden-state matching depth during knowledge transfer.}
\label{tab:hid_match}
\begin{tabular}{lccc}
\toprule
Configuration & AUROC & AUPRC & Acc.\ (\%) \\
\midrule
Layers 3,5,7 & $.989$ & $.977$ & $93.97$ \\
$\mathcal{L}_{\text{hid}}{=}0$ (final only) & $.953$ & $.922$ & $90.95$ \\
\textbf{All layers} & $\mathbf{.994}$ & $\mathbf{.988}$ & $\mathbf{96.08}$ \\
No-teacher (baseline) & $.963$ & $.937$ & $91.64$ \\
\bottomrule
\end{tabular}
\end{table}

Supervising all hidden layers yields the best performance (\cref{tab:hid_match}), and removing hidden-state matching entirely degrades metrics \emph{below} the no-teacher baseline, indicating multi-depth guidance is essential for stable cross-modal transfer. This suggests intermediate representations carry complementary structural information about the privileged modality that final-layer-only transfer cannot capture.

\subsection{Shared--Private Capacity}\label{subsec:private_mask}

\begin{table}[t]
\centering
\small
\caption{Private-mask ratio controlling shared vs.\ private capacity.}
\label{tab:private_mask}
\begin{tabular}{lccc}
\toprule
Ratio & AUROC & AUPRC & Acc.\ (\%) \\
\midrule
0 (all shared) & $.959$ & $.927$ & $70.41$ \\
0.2 & $.980$ & $.966$ & $\mathbf{94.46}$ \\
0.5 (default) & $\mathbf{.989}$ & $\mathbf{.977}$ & $93.97$ \\
0.8 & $.975$ & $.958$ & $93.76$ \\
1 (no transfer) & $.945$ & $.906$ & $70.41$ \\
\bottomrule
\end{tabular}
\end{table}

Both extremes hurt (\cref{tab:private_mask}): no private capacity (ratio$=$0) eliminates modality-specific features needed for reconstruction; no shared capacity (ratio$=$1) removes the pathway for privileged transfer. A balanced split (0.5) maximizes ranking metrics, validating the shared--private decomposition as essential rather than incidental.

\section{Discussion: Generalization to Broader Embodied Sensing}\label{sec:discussion}

While our experiments use wearable physiological signals, the core insights of PULSE address challenges shared across embodied multi-sensory systems:

\paragraph{Why EDA is a uniquely informative teacher.}
EDA's privileged status for stress monitoring is grounded in physiology, not convenience.
Eccrine sweat glands are innervated exclusively by the sympathetic branch of the autonomic nervous system; changes in skin conductance therefore provide a direct, peripheral readout of central sympathetic activation during acute stress~\cite{critchley2002brain,boucsein2012EDA}.
No other wrist-worn modality offers this directness: ECG reflects cardiac autonomic control (both sympathetic and parasympathetic), BVP captures peripheral vasomotion modulated by multiple regulatory loops, and accelerometry and temperature are only indirectly correlated with arousal.
This physiological asymmetry is precisely what makes EDA an effective teacher: it carries information about stress-related autonomic states that cheaper sensors cannot directly observe, creating a natural supervisory gradient for knowledge transfer.

\paragraph{What wearable physiological sensors enable.}
Wearable biosensors occupy a unique niche in embodied sensing that distinguishes them from conventional vision or audio modalities.
They capture internal autonomic and cardiovascular states, processes with no direct visual manifestation, enabling monitoring during sleep, under clothing, and in private settings where cameras are socially unacceptable or legally prohibited.
Their continuous, longitudinal operation (days to weeks on a single charge) supports applications such as chronic stress tracking, seizure detection, and circadian rhythm analysis that episodic visual sensing cannot address.
PULSE is designed for precisely this deployment context: rich laboratory sensors (EDA) are available only during controlled data collection, while cheap, always-on wearable sensors (ECG, BVP, ACC, TEMP) must carry the full inference burden in unconstrained daily life.

\paragraph{Sensor asymmetry is ubiquitous.}
Dense tactile arrays on robotic grippers provide rich contact information during lab data collection but are too fragile or expensive for field deployment, where only sparse force sensors and IMUs are available.
High-fidelity bioelectrical sensors (EDA, EMG, EEG) capture neural and autonomic processes during controlled human studies but are impractical during dexterous manipulation, where electrodes interfere with grasping.
In XR, research headsets carry a full complement of physiological sensors, but consumer hardware retains only optical heart-rate and inertial measurement.
In all cases, PULSE's formulation applies: freeze a teacher trained on the rich modality and distill into deployment-grade students.

\paragraph{Shared--private decomposition as a general principle.}
The decomposition into modality-invariant and modality-specific subspaces is motivated by a general observation: not all information in one sensor should be forced to match another.
In the wearable stress domain, the shared subspace captures the common autonomic arousal signature that EDA, ECG, and BVP all reflect (albeit with different fidelity), while the private subspace preserves modality-specific morphology: QRS complex shape and R-R interval dynamics in ECG, pulse wave amplitude and transit-time features in BVP, and motion intensity patterns in accelerometry.
Forcing these distinct signal structures into a single aligned space destroys information that downstream tasks may need (e.g., QRS morphology for arrhythmia screening alongside stress classification).
More broadly, in tactile-to-IMU transfer, the spatially-resolved pressure distribution (private to tactile) differs fundamentally from the whole-body dynamics captured by the IMU; only the shared contact-event structure should be aligned. Our ablations confirm that collapsing private capacity (ratio$=$0) or shared capacity (ratio$=$1) degrades performance.

\paragraph{Reconstruction as a collapse safeguard.}
We demonstrate that knowledge distillation without an auxiliary reconstruction objective leads to representational collapse. When we train without reconstruction (KD-only), the shared embeddings collapse toward a constant vector across modalities: the mean pairwise cosine $\approx 1.0$ for ECG/BVP/ACC (TEMP 0.998), and the feature variance is near-zero, indicating almost no dispersion. After adding the reconstruction objective, feature variance returns by several orders of magnitude, and mean pairwise cosine drops to 0.027--0.137, showing that samples are no longer trivially aligned (\cref{fig:reconstruction_effect}). This pattern holds consistently across modalities, supporting our claim that reconstruction acts as an information-preserving regularizer: it counteracts the KD objective's tendency to minimize alignment by shrinking representation variance, thereby preventing representational collapse and yielding meaningful geometry for downstream LOSO generalization.

\begin{figure}[t]
\centering

\begin{minipage}{\columnwidth} 
\centering
\small 
\setlength{\tabcolsep}{4pt} 
\resizebox{\columnwidth}{!}{
\begin{tabular}{lcc}
\toprule
Modality & mean pairwise cosine & mean feature variance \\
\midrule
ECG  & 1.000 & 2.57e--05 \\
BVP  & 1.000 & 6.26e--05 \\
ACC  & 1.000 & 7.53e--05 \\
TEMP & 0.998 & 7.96e--04 \\
\bottomrule
\end{tabular}
}
\caption*{\small Without reconstruction}
\end{minipage}

\vspace{0.6em} 

\begin{minipage}{\columnwidth} 
\centering
\small 
\setlength{\tabcolsep}{4pt} 
\resizebox{\columnwidth}{!}{
\begin{tabular}{lcc}
\toprule
Modality & mean pairwise cosine & mean feature variance \\
\midrule
ECG  & 0.070  & 2.52e--02 \\
BVP  & 0.0645 & 1.51e--02 \\
ACC  & 0.137  & 9.20e--02 \\
TEMP & 0.0273 & 2.06e--03 \\
\bottomrule
\end{tabular}
}
\caption*{\small With reconstruction}
\end{minipage}

\caption{Effect of adding reconstruction during KD. Without reconstruction, shared embeddings collapse (cosine $\approx 1.0$, near-zero variance). Adding reconstruction restores variance by several orders of magnitude, preventing collapse and enabling meaningful shared geometry.}
\label{fig:reconstruction_effect}
\end{figure}

\paragraph{Multi-depth distillation.}
Our ablations in Table \ref{tab:hid_match} show all-layer hidden-state matching substantially outperforms final-only transfer. This suggests that intermediate representations carry complementary structural information about the privileged modality, a finding likely transferable to other encoder architectures processing heterogeneous sensor streams.

\paragraph{Evidence that shared embeddings encode privileged EDA knowledge.}
Several lines of existing evidence support the claim that the shared subspace genuinely captures EDA-derived stress information rather than merely learning generic modality-invariant features.
First, the performance gap between PULSE and the no-teacher baseline (which uses the same architecture, pretraining, and shared--private decomposition but without EDA-guided distillation) is entirely attributable to the frozen EDA teacher: the only difference between configurations A and C (\cref{tab:finetune}) is the presence of knowledge transfer from EDA, yet AUROC improves from 0.963 to 0.994.
Second, the collapse analysis in \cref{fig:reconstruction_effect} shows that with reconstruction, shared embeddings develop structured geometry (mean pairwise cosine 0.027--0.137, variance restored by three orders of magnitude), and this geometry is shaped by the EDA teacher's hidden states at every encoder layer.
Third, the modality-agnostic ablation (\cref{tab:app_ecg_teacher}, \cref{app:ecg_teacher}) shows that replacing EDA with ECG as teacher produces smaller gains, consistent with the interpretation that teacher informativeness, specifically EDA's direct coupling to sympathetic arousal, determines how much privileged knowledge is encoded in the shared subspace.
We acknowledge that a more direct validation, such as probing classifiers on the shared embeddings or t-SNE visualization comparing shared representations with and without EDA-guided transfer, would further strengthen this claim; we leave such analyses to future work.

\section{Conclusion}\label{sec:conclusion}

We present PULSE, a privileged knowledge transfer framework that addresses the sensor-asymmetry problem in multi-sensory embodied systems.
By decomposing student representations into shared and private embeddings, aligning only the shared subspace with a frozen privileged-sensor teacher, and using reconstruction to prevent collapse, PULSE transfers information from rich but impractical sensors into deployment-ready modalities.
On WESAD, PULSE achieves state-of-the-art stress detection without EDA at inference, matching if not surpassing the full-sensor model.
We further validate the framework on the PhysioNet STRESS dataset, demonstrating generalization across different populations and stressor protocols.

The framework's design principles, namely shared--private decomposition, multi-depth distillation, hinge-based alignment, and reconstruction-based regularization, are sensor- and task-agnostic.
We envision applications in tactile-to-inertial transfer for robotic manipulation, bioelectrical-to-optical transfer for consumer wearables, and multi-sensory XR systems where privileged data is available only during development.
Future work will validate PULSE on additional datasets and sensing domains, explore multi-teacher ensembles, study cross-dataset transfer to quantify domain generalization, and provide direct interpretability analyses (e.g., probing classifiers, t-SNE visualization) of the shared embedding space.

\newpage
{\small
\bibliographystyle{ieeenat_fullname}
\bibliography{main}
}

\newpage
\appendix
\section{Comparison with Existing Work}\label{apd:comparison}

\begin{table*}[htbp]
\centering
\footnotesize
\setlength{\tabcolsep}{4pt}
\renewcommand{\arraystretch}{1.12}
\caption{Selected WESAD results from prior work, augmented with our results. Metrics are reported as in the original papers and are \emph{not} directly comparable across differing task setups (binary vs.\ 3/4-class) and evaluation protocols.}
\label{tab:wesad_compare}
\begin{tabular}{@{}%
  >{\RaggedRight\arraybackslash}p{0.31\textwidth}%
  >{\RaggedRight\arraybackslash}p{0.31\textwidth}%
  >{\RaggedRight\arraybackslash}p{0.20\textwidth}%
  >{\centering\arraybackslash}p{0.09\textwidth}%
  >{\centering\arraybackslash}p{0.07\textwidth}@{}}
\toprule
\textbf{Paper (Year)} & \textbf{Test-time sensors} & \textbf{Eval protocol} & \textbf{Accuracy (\%)} & \textbf{AUROC} \\
\midrule
\textbf{Ours (PULSE, 2025)} & ECG + BVP + ACC + TEMP & LOSO, binary & $93.97 \pm 5.77$ & $0.989 \pm 0.017$ \\
\textbf{Ours (Symmetric alignment, 2025)} & ECG + BVP + ACC + TEMP & LOSO, binary & $88.83 \pm 6.24$ & $0.972 \pm 0.031$ \\
\textbf{Ours (No-EDA baseline, 2025)} & ECG + BVP + ACC + TEMP & LOSO, binary & $91.64 \pm 6.61$ & $0.963 \pm 0.050$ \\
\textbf{Ours (Full-sensor baseline, 2025)} & ECG + BVP + ACC + TEMP + EDA & LOSO, binary & $90.74 \pm 5.58$ & $0.983 \pm 0.028$ \\
\midrule
\cite{Prajod_2022} & ECG only & LOSO, binary & 90.8 & N/A \\
\cite{Behinaein_2021} & ECG only & LOSO, binary$^{\dagger}$ & 91.1 & N/A \\
\cite{karan2021timeseries} & ECG only & LOSO, binary & 88.7 & N/A \\
\cite{liakopoulos2021cnn} & ECG only & LOSO, binary & 82.4 & N/A \\
\cite{oliver2025crossmodality} & ECG / BVP / EDA / RESP / TEMP (single-modality) & Subject-dependent (random 85{:}15) & up to 99.95 (validation; multiclass) & N/A \\
\cite{Yang_2025} & EDA + ACC + BVP + RESP + TEMP & LOSO, 3-class & 90.96 & N/A \\
\cite{Singh_2024} & ECG + EDA + RESP (+TEMP/ACC) & LOSO, 3-class & 90.45 & N/A \\
\cite{Ghosh_2022} & EDA + ECG + ACC + RESP + TEMP & LOSO, 4-class & 94.77 & N/A \\
\bottomrule
\end{tabular}
\end{table*}

We show a comparison of our results with some existing work in \cref{tab:wesad_compare}. Most WESAD reports either (i) rely on a single sensor (often ECG) or (ii) include EDA at inference, and many use task definitions and splits that differ from ours (e.g., multi-class vs.\ binary; subject-dependent vs.\ LOSO). Consequently, these numbers provide useful context but are not like-for-like baselines for our setting where EDA is absent at test time. Behinaein et al \cite{Behinaein_2021} reports 91.1\% accuracy only after fine-tuning/calibration using 10\% of the held-out test subject; the zero-calibration LOSO accuracy is 80.4\%.

\section{Additional Ablation Studies}\label{app:ablation}

We report additional hyperparameter sweeps and secondary ablations. The collapse analysis supporting the reconstruction regularizer is in \cref{fig:reconstruction_effect} of the main text. Unless noted, metrics are mean~$\pm$~std across 15 LOSO folds.

\subsection{Modality Dropout at Test Time}\label{app:modality_drop_test}

We assess the sensitivity of PULSE to missing test-time modalities. Results are obtained by removing signal modalities during finetuning and inference. Relative to the full setting (ECG+BVP+ACC+TEMP), removing TEMP yields small changes. Removing ACC in addition (ECG+BVP) produces the largest additional drop. Going to ECG only leads to a further but modest decline.

Variability increases as modalities are removed (e.g., AUPRC std: $0.033 \rightarrow 0.036 \rightarrow 0.088 \rightarrow 0.117$), indicating reduced stability under stronger input constraints. Overall, PULSE degrades gracefully and maintains strong performance even with ECG only, consistent with effective transfer of privileged-sensor structure into each student.

\begin{table*}[htbp]
\centering
\small
\setlength{\tabcolsep}{8pt}
\renewcommand{\arraystretch}{1.1}
\caption{Ablations on test-time input availability for PULSE. Reported means and standard deviations are computed across folds.}
\label{tab:app_modality_drop_test}
\begin{tabular}{llccc}
\toprule
\textbf{Name} & \textbf{Test-time inputs} & \textbf{AUROC} & \textbf{AUPRC} & \textbf{Accuracy (\%)} \\
\midrule
PULSE (full)        & ECG + BVP + ACC + TEMP & $\mathbf{0.989 \pm 0.017}$ & $\mathbf{0.977 \pm 0.033}$ & $\mathbf{93.97 \pm 5.77}$ \\
PULSE ($-$TEMP)       & ECG + BVP + ACC        & $0.985 \pm 0.021$ & $0.972 \pm 0.036$ & $90.85 \pm 6.53$ \\
PULSE ($-$TEMP, $-$ACC) & ECG + BVP              & $0.966 \pm 0.072$ & $0.948 \pm 0.088$ & $87.74 \pm 8.53$ \\
PULSE (ECG only)    & ECG                    & $0.960 \pm 0.082$ & $0.936 \pm 0.117$ & $86.22 \pm 10.37$ \\
\bottomrule
\end{tabular}
\end{table*}

\subsection{Modality Dropout During Pretraining}\label{app:modality_drop_pre}

\begin{table*}[htbp]
\centering
\small
\setlength{\tabcolsep}{10pt}
\renewcommand{\arraystretch}{1.1}
\caption{Effect of removing modalities during pretraining. Full-sensor pretraining yields the strongest downstream performance.}
\label{tab:app_modality_drop_pre}
\begin{tabular}{lccc}
\toprule
\textbf{Setup} & \textbf{AUROC} & \textbf{AUPRC} & \textbf{Accuracy}\\
\midrule
PULSE (ECG + BVP + ACC + TEMP) & $\mathbf{0.989 \pm 0.017}$ & $\mathbf{0.977 \pm 0.033}$ & $\mathbf{93.97 \pm 5.77}$\,\% \\
ECG + BVP + ACC                 & $0.949 \pm 0.084$ & $0.923 \pm 0.117$ & $86.12 \pm 8.22$\,\% \\
ECG + BVP                       & $0.973 \pm 0.048$ & $0.942 \pm 0.112$ & $90.44 \pm 9.92$\,\% \\
Only ECG                        & $0.960 \pm 0.077$ & $0.927 \pm 0.137$ & $75.74 \pm 10.81$\,\% \\
\bottomrule
\end{tabular}
\end{table*}

Pretraining with all four cheap sensors yields the best results; dropping modalities degrades AUPRC and accuracy most sharply, indicating reduced transfer of privileged EDA structure. Compared with \cref{tab:app_modality_drop_test}, removing modalities during pretraining degrades performance more than removing them at test time, underscoring the benefits of multimodal representation learning over single-modality training.

\subsection{Modality-Agnostic Transfer: ECG as Teacher}\label{app:ecg_teacher}

To evaluate whether privileged knowledge transfer in PULSE depends specifically on EDA or can generalize to other physiological modalities, we conducted an ablation using \textbf{ECG as the privileged teacher}. In this setup, the pretrained ECG encoder is frozen and used to distill knowledge into the remaining deployable modalities (BVP, ACC, and TEMP) through the same shared/private alignment and hidden-state matching objectives. \cref{tab:app_ecg_teacher} reports the results under the LOSO protocol.

Using ECG as the teacher yields moderate improvements in AUROC and AUPRC compared to the no-KD counterpart, while accuracy remains comparable within variance. This demonstrates that the PULSE framework is \emph{modality-agnostic}: privileged distillation benefits are not limited to EDA, though the magnitude of gain depends on the informativeness of the teacher modality. EDA remains the strongest privileged signal for stress-related supervision due to its direct coupling with sympathetic arousal, yet ECG-based distillation still provides measurable transfer, confirming that PULSE can flexibly leverage alternative privileged channels when EDA is unavailable.

\begin{table*}[t]
\centering
\caption{ECG as frozen teacher distilled to BVP/ACC/TEMP students.}
\label{tab:app_ecg_teacher}
\begin{tabular}{lccc}
\toprule
Setup & AUROC & AUPRC & Accuracy (\%) \\
\midrule
ECG teacher $\rightarrow$ (BVP, ACC, TEMP) & $\mathbf{0.956 \pm 0.067}$ & $\mathbf{0.930 \pm 0.085}$ & $86.63 \pm 5.86$ \\
(BVP, ACC, TEMP) w/o KD                    & $0.949 \pm 0.058$ & $0.902 \pm 0.102$ & $\mathbf{87.06 \pm 8.77}$ \\
\bottomrule
\end{tabular}
\end{table*}

\subsection{Hinge-Loss Margin}\label{app:hinge_margin}

\begin{table*}[htbp]
\centering
\small
\setlength{\tabcolsep}{10pt}
\renewcommand{\arraystretch}{1.1}
\caption{Effect of hinge-loss margin for shared-embedding alignment. A small positive margin (0.2) is optimal; too large a margin degrades performance, likely by over-separating positive pairs and harming transfer.}
\label{tab:app_margin}
\begin{tabular}{lccc}
\toprule
\textbf{Margin} & \textbf{AUROC} & \textbf{AUPRC} & \textbf{Accuracy} \\
\midrule
0.0            & $0.972 \pm 0.039$ & $0.954 \pm 0.060$ & $70.41 \pm 8.60$\,\% \\
0.2 (default)  & $\mathbf{0.989 \pm 0.017}$ & $\mathbf{0.977 \pm 0.033}$ & $\mathbf{93.97 \pm 5.77}$\,\% \\
0.4            & $0.977 \pm 0.051$ & $0.966 \pm 0.071$ & $89.08 \pm 13.39$\,\% \\
0.6            & $0.983 \pm 0.052$ & $0.975 \pm 0.070$ & $77.57 \pm 8.85$\,\% \\
0.8            & $0.957 \pm 0.077$ & $0.938 \pm 0.112$ & $70.41 \pm 8.60$\,\% \\
1.0            & $0.969 \pm 0.070$ & $0.933 \pm 0.160$ & $76.43 \pm 8.69$\,\% \\
\bottomrule
\end{tabular}
\end{table*}

A modest margin (0.2) provides the strongest signal for aligning student shared embeddings with the EDA teacher; margins that are too small (0) or too large ($\geq 0.6$) underperform.

\subsection{Shared--Private Capacity (Full Sweep)}\label{app:private_mask}

\begin{table*}[htbp]
\centering
\small
\setlength{\tabcolsep}{10pt}
\renewcommand{\arraystretch}{1.1}
\caption{Varying the private-mask ratio that reserves capacity for modality-specific features. A balanced split (0.5) maximizes AUROC/AUPRC, while a slightly smaller private region (0.2) peaks accuracy. Removing knowledge transfer (ratio~1) hurts all metrics.}
\label{tab:app_private_mask}
\begin{tabular}{lccc}
\toprule
\textbf{Private mask ratio} & \textbf{AUROC} & \textbf{AUPRC} & \textbf{Accuracy} \\
\midrule
0                            & $0.959 \pm 0.062$ & $0.927 \pm 0.100$ & $70.41 \pm 8.60$\,\% \\
0.2                          & $0.980 \pm 0.044$ & $0.966 \pm 0.076$ & $\mathbf{94.46 \pm 7.78}$\,\% \\
0.4                          & $0.919 \pm 0.116$ & $0.894 \pm 0.148$ & $79.51 \pm 13.92$\,\% \\
0.5 (default)                & $\mathbf{0.989 \pm 0.017}$ & $\mathbf{0.977 \pm 0.033}$ & $93.97 \pm 5.77$\,\% \\
0.6                          & $0.967 \pm 0.072$ & $0.956 \pm 0.084$ & $92.62 \pm 9.89$\,\% \\
0.8                          & $0.975 \pm 0.047$ & $0.958 \pm 0.082$ & $93.76 \pm 5.96$\,\% \\
1 (no knowledge transfer)    & $0.945 \pm 0.078$ & $0.906 \pm 0.127$ & $70.41 \pm 8.60$\,\% \\
\bottomrule
\end{tabular}
\end{table*}

Both insufficient private capacity (ratio~0: no private embeddings to encode modality-specific features) and excessive private capacity (ratio~1: no shared embeddings for alignment and knowledge transfer) hurt performance. A \emph{balanced} split between shared and private capacity (0.5) most effectively supports privileged EDA transfer, while a smaller private fraction (0.2) trades a slight drop in AUROC/AUPRC for the highest accuracy.

\subsection{Fusion Strategy}\label{app:fusion}

We examined whether learning per-modality fusion weights could improve performance compared to static averaging of modality embeddings. In the \textbf{adaptive fusion} variant, a small trainable gating module predicts fusion weights for each modality during both knowledge distillation and finetuning, initialized to uniform values and updated end-to-end. The default PULSE configuration uses a \textbf{static average} of shared embeddings across modalities.

\begin{table*}[t]
\centering
\caption{Fusion strategy during finetuning.}
\label{tab:app_fusion}
\begin{tabular}{lccc}
\toprule
Model & AUROC & AUPRC & Accuracy (\%) \\
\midrule
PULSE (static fusion; default) & $\mathbf{0.989 \pm 0.017}$ & $\mathbf{0.977 \pm 0.033}$ & $\mathbf{93.97 \pm 5.77}$ \\
PULSE (adaptive fusion)        & $0.967 \pm 0.046$ & $0.951 \pm 0.064$ & $90.91 \pm 7.35$ \\
\bottomrule
\end{tabular}
\end{table*}

Adaptive fusion offered no consistent improvement and slightly degraded performance. We attribute this to the high inter-subject variability under LOSO evaluation: fusion weights learned on training subjects do not necessarily generalize to unseen participants. Static averaging enforces a more stable modality combination and avoids overfitting to subject-specific patterns.

\subsection{Alignment Loss}\label{app:align_loss}

To quantify the contribution of the cross-modal alignment objective, we conducted an ablation removing $\mathcal{L}_{\text{align}}$ during pretraining. In this setting, each modality's encoder is trained solely via its reconstruction objective (MAE) without enforcing consistency across shared embeddings.

\begin{table*}[htbp]
\centering
\caption{Ablation on alignment loss $\mathcal{L}_{\text{align}}$ during pretraining. Removing alignment substantially degrades performance, indicating that shared-space regularization is critical for multimodal representation learning.}
\label{tab:app_align_loss}
\begin{tabular}{lccc}
\toprule
Model & AUROC & AUPRC & Accuracy (\%) \\
\midrule
no-EDA baseline (with alignment loss)   & $\mathbf{0.963 \pm 0.050}$ & $\mathbf{0.937 \pm 0.101}$ & $\mathbf{91.64 \pm 6.61}$ \\
no-EDA baseline (without alignment loss) & $0.945 \pm 0.078$ & $0.906 \pm 0.127$ & $70.41 \pm 8.60$ \\
\bottomrule
\end{tabular}
\end{table*}

Removing the alignment loss leads to a sharp drop in both ranking metrics and accuracy, confirming that explicit cross-modal regularization is essential for learning a coherent shared latent space.

\subsection{Hidden-State Matching (Full Results)}\label{app:hid_match}

To assess the effect of hidden-state supervision during knowledge distillation, we performed ablations over which layers of the student were matched to the frozen EDA teacher. We compare three configurations: (i) removing hidden-state matching entirely ($\mathcal{L}_{\text{hid}} = 0$), equivalent to final-representation-only transfer; (ii) matching only intermediate layers (3, 5, 7); and (iii) matching all hidden layers.

\begin{table*}[t]
\centering
\caption{Effect of hidden-state matching during KD. Full results with standard deviations.}
\label{tab:app_hid_match}
\begin{tabular}{lccc}
\toprule
Model & AUROC & AUPRC & Accuracy (\%) \\
\midrule
PULSE (match layers 3,5,7) & $0.989 \pm 0.017$ & $0.977 \pm 0.033$ & $93.97 \pm 5.77$ \\
PULSE ($\mathcal{L}_{\text{hid}}{=}0$; final-only) & $0.953 \pm 0.096$ & $0.922 \pm 0.155$ & $90.95 \pm 9.73$ \\
\textbf{PULSE (match all layers)} & $\mathbf{0.994 \pm 0.011}$ & $\mathbf{0.988 \pm 0.022}$ & $\mathbf{96.08 \pm 4.52}$ \\
no-EDA baseline (no KD)    & $0.963 \pm 0.050$ & $0.937 \pm 0.101$ & $91.64 \pm 6.61$ \\
\bottomrule
\end{tabular}
\end{table*}

Removing hidden-state matching degrades all metrics relative to both PULSE and the no-EDA baseline, indicating that intermediate feature supervision is critical for transferring sympathetic-arousal structure. Supervising \emph{all} layers yields the highest AUROC, AUPRC, and accuracy, suggesting that multi-depth guidance stabilizes optimization and enables richer cross-modal transfer.

\section{Why Hinge Loss (vs.\ Anchored Contrastive)}\label{app:hinge_vs_contrastive}

\begin{figure*}[htbp]
    \centering
    \includegraphics[width=0.5\linewidth]{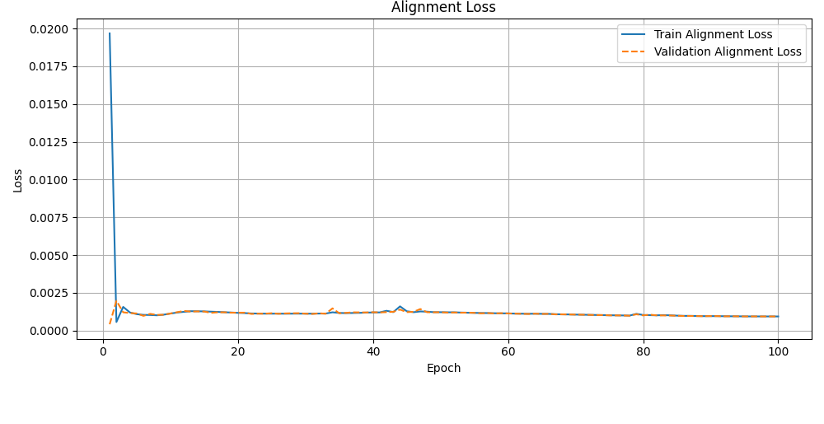}
    \caption{Learning curve showing anchored contrastive loss in early experiments.}
    \label{fig:old_loss}
\end{figure*}

During pretraining we first tried an \textbf{anchored contrastive} objective: each modality's shared embeddings were aligned to ECG (the ``anchor'') via cosine similarity. In practice, the alignment loss collapsed to ${\sim}0$ within the first few iterations and then remained nearly constant, providing little training signal thereafter, as shown in \cref{fig:old_loss}. This early collapse motivated switching to a \textbf{hinge-based alignment} with a small positive margin, which maintained non-trivial gradients throughout training.

\section{Training Dynamics}\label{app:training_dynamics}

\begin{figure*}[htbp]
    \centering
    \includegraphics[width=1.0\linewidth]{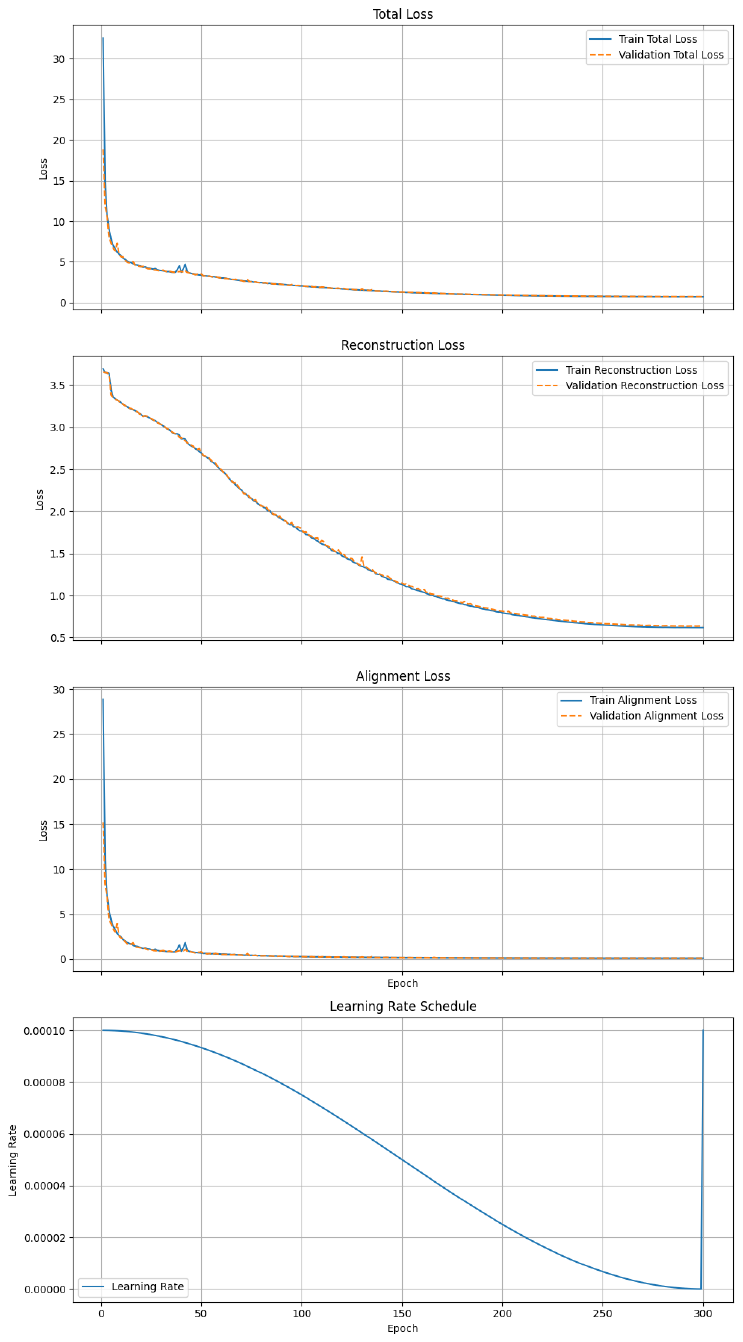}
    \caption{Convergence behavior in pretraining.}
    \label{fig:pretrain_loss}
\end{figure*}

\begin{figure*}[htbp]
    \centering
    \includegraphics[width=1\linewidth]{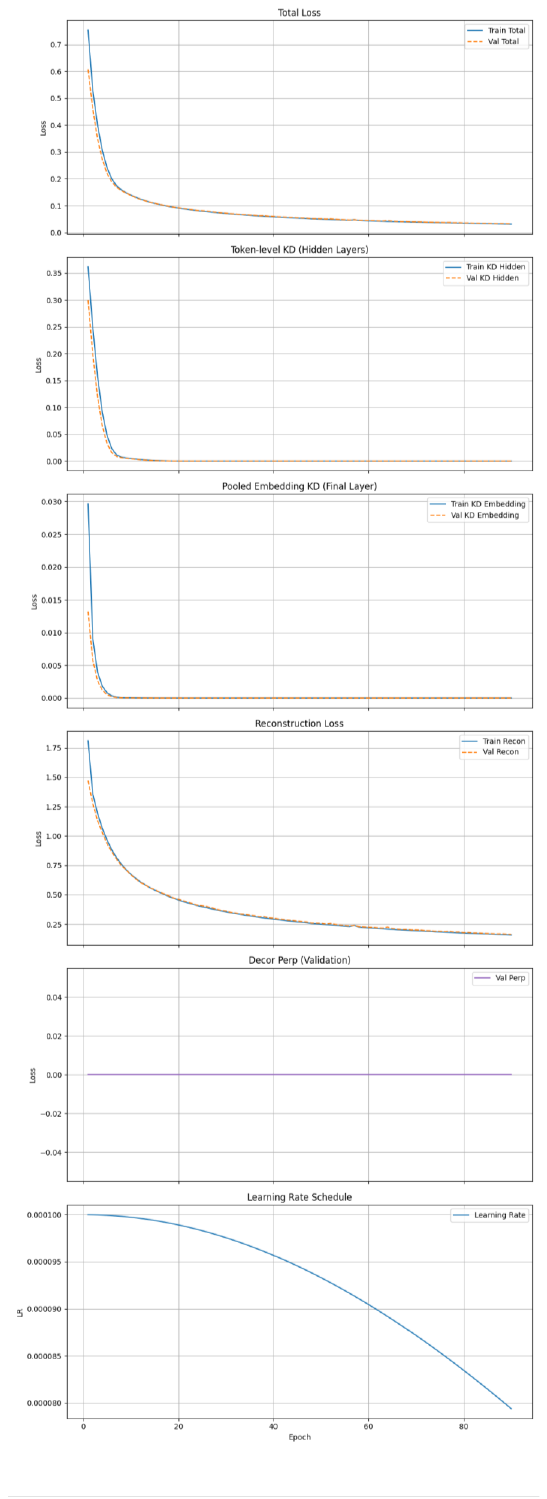}
    \caption{Convergence behavior in knowledge transfer.}
    \label{fig:kd_loss}
\end{figure*}

\begin{figure*}[htbp]
    \centering
    \includegraphics[width=1\linewidth]{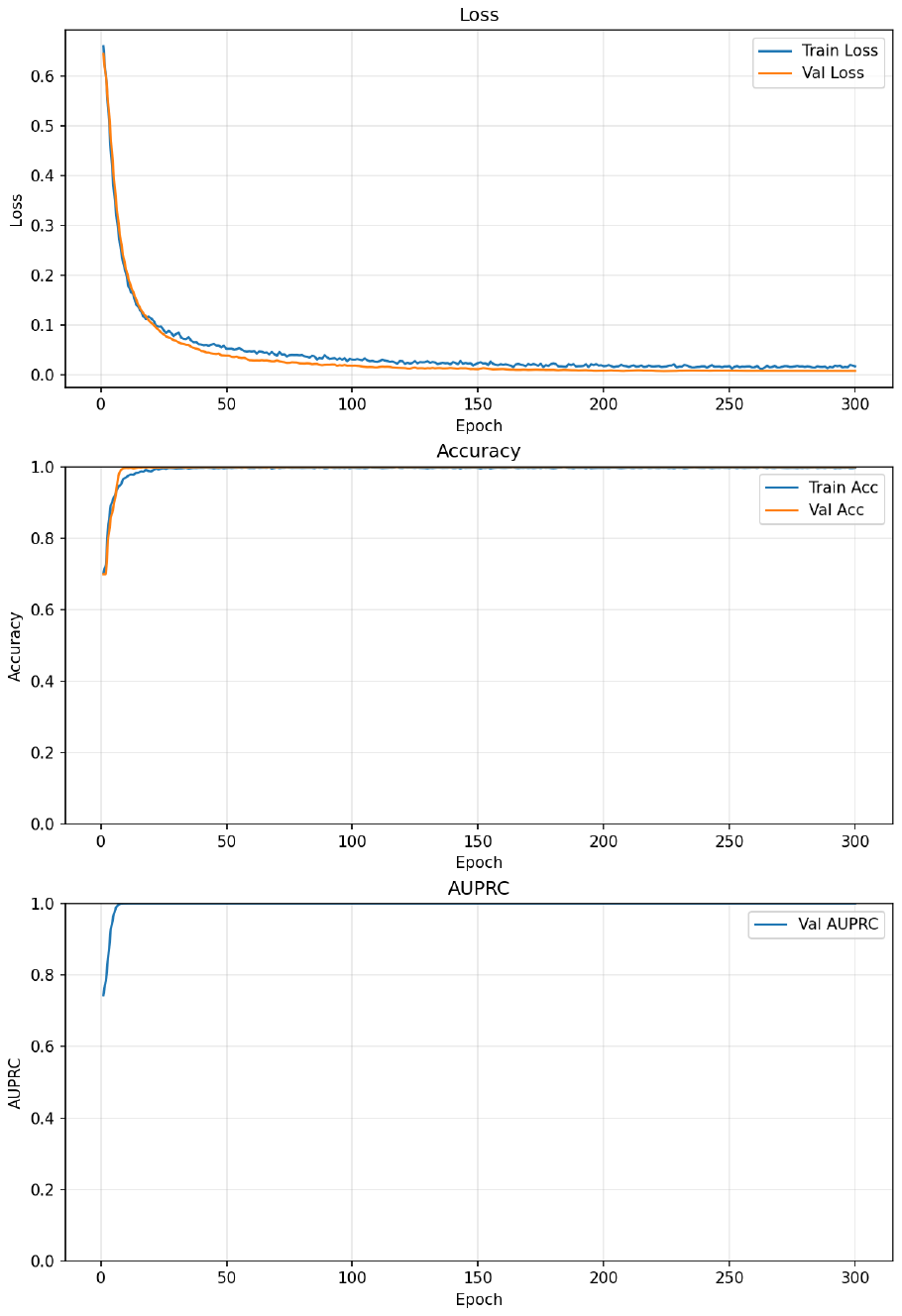}
    \caption{Convergence behavior in finetuning.}
    \label{fig:finetune_loss}
\end{figure*}

In \cref{fig:pretrain_loss,fig:kd_loss,fig:finetune_loss}, we show the convergence behavior in different stages of training (pretraining, knowledge transfer, and finetuning).

\end{document}